# Coherent Processing of a Qubit Using One Squeezed State

Allan Tameshtit

allan.tameshtit@utoronto.ca

**Abstract:** In a departure from most work in quantum information utilizing Gaussian states, we use a single such state to represent a qubit and model environmental noise with a class of quadratic dissipative equations. A benefit of this single Gaussian representation is that with one deconvolution, we can eliminate noise. In this deconvolution picture, a basis of squeezed states evolves to another basis of such states. One of the limitations of our approach is that noise is eliminated only at a privileged time. We suggest that this limitation may actually be used advantageously to send information securely: the privileged time is only known to the sender and the receiver, and any intruder accessing the information at any other time encounters noisy data.

## 1. Introduction

In quantum computing, errors arising from quantum noise may be treated using quantum error-correcting codes, such as the Shor code [1]. The basic idea is to convert a qubit that one is trying to shield from noise to a higher dimensional state containing redundant information. After being subjected to noise, a recovery procedure is implemented to rehabilitate the state. Departing from this approach, the work presented below exploits a unique factorization of quadratic dissipative equations to leave squeezed states coherent and accessible to quantum processing after a deconvolution at a privileged time $t^*$. This is accomplished by using a single Gaussian state to represent a qubit. Quantum noise is thus eliminated without having to increase the dimensions of the system under study, albeit at the expense of having to perform the deconvolution at $t^*$. We suggest that this limitation may be used advantageously to help send information securely. Only authorized persons would be apprised of $t^*$. An intruder accessing the information at any other time encounters noisy data.

Gaussian states have been widely used for continuous variable quantum information processing (see [2] for a review). Experimental examples include teleporting a single coherent state [3] and investigating transmitted states of light represented by a mixture of two coherent states [4,5]. Theoretically, a superposition of Gaussian states has been used to represent states of a two-dimensional code space [6]. Another application of Gaussian states employs a superposition of two substantially non-overlapping coherent or squeezed states as the computational basis to represent a qubit [7,8,9]. Because of the small overlap, this superposition is prone to decoherence, a bane of information processing. Error correction codes for continuous variables have been developed to reduce the deleterious noise effects [2,4,5,6,10,11]. Here, we abandon superpositions of Gaussians and instead utilize a single squeezed state to represent a qubit, reversing decoherence with an impurity filter. We encode information using the phase space position of the state, as in quantum key distribution protocols [2] that use coherent [12] or squeezed states [13,14]. The application of the impurity filter amounts to performing a Wigner function deconvolution, which is in line with the general use of deconvolutions to eliminate noise, such as instrumental quantum noise [15,16]. In quantum homodyne tomography, for example, deconvolutions have been used to reconstruct Wigner functions from noisy data [17].

**2. Review of Quadratic Dissipative Equations**

As our testing ground for handling quantum noise, we will utilize general quadratic dissipative equations (QDEs), describing the evolution of an oscillator subject to fluctuations and dissipation. QDEs are a class of master equations with time dependent coefficients, also known as non-autonomous equations. In this section, we will briefly review such equations and mention a few well-known master equations that are examples of QDEs. Some lessons from these examples will also guide us when we attempt to eliminate noise in general QDEs. The reader interested in more details of QDEs can consult Reference [18].

A QDE is the following equation governing the evolution of the density operator, $\rho$:

$$\frac{d}{dt}\rho(t) = \frac{1}{\hbar i}[H_s(t),\rho(t)] - k_1(t)\{q,\rho(t),q\} - k_2(t)\{p,\rho(t),p\} + k_3(t)\{p,\rho(t),q\} + k_4(t)\{q,\rho(t),p\} \quad (1)$$

where the system Hamiltonian is given by $H_s(t) \equiv b_{11}(t)q^2 + b_{12}(t)qp + b_{21}(t)pq + b_{22}(t)p^2$, with $b_{11}$, $b_{12} = b_{21}$, $b_{22}$, $k_1$ and $k_2$ being real functions, $k_3$ being a complex function, $k_4 = k_3^*$ and $\{A,\rho,B\} \equiv BA^\dagger\rho + \rho BA^\dagger - 2A^\dagger\rho B$. As motivation for studying QDEs, we note that most one-dimensional quantum master equations describing harmonic motion in the literature, both exact and approximate, are of this form. For example, a subset of QDEs that are identified with Brownian motion has been solved exactly as a Wigner function convolution [19,20,21]. Another well known, autonomous example of a QDE, is the rotating wave optical master equation for an oscillator, which, when written as it usually is in terms of lowering and raising operators, is given in the interaction picture by (see, e.g., [22] and references therein)

$$\frac{d\tilde{\rho}^I}{dt} = -\frac{\tilde{\gamma}}{2}[(\bar{n}+1)\{a^\dagger,\rho^I,a^\dagger\} + \bar{n}\{a,\rho^I,a\}] \quad (2)$$

where $\tilde{\gamma}$ is a measure of the coupling strength between the oscillator and its environment, and $\bar{n}$ is the average number of oscillator quanta in a state of thermal equilibrium.

Although Equation (2) is one of the simplest QDEs, certain features of this equation foreshadow our approach for reducing quantum noise in more general cases. While at all temperatures the pure states that maximize the initial rate of change of the purity $Tr\rho^2$ predicted by Equation (2) are the ordinary coherent states [23,24], it is only at absolute zero, when $\bar{n} = 0$ and the $\{a,\rho^I,a\}$ term disappears, that these coherent states—and only these states—remain pure as they evolve [24]. The key to this result is that at absolute zero, only one dissipative operator of the form $\{a^\dagger,\rho^I,a^\dagger\}$ remains in Equation (2), an idea that we will use below when investigating more general QDEs.

With the goal of determining when states can remain pure, Hasse has examined other autonomous master equations [25]. In this last reference, non-linear Schrodinger equations that describe the evolution of the corresponding wave function were formulated.

These examples of decoherence-free evolution motivate us to inquire about similar behaviour in more general, non-autonomous QDEs. We will see that in the general case, we have to content ourselves with less. In particular, while it is not generally possible for a coherent state to remain coherent for all times, below we will show that by making use of a deconvolution, we can ensure that a squeezed state evolves to another squeezed state at a privileged time.

Before leaving this section, we will provide the solution of Equation (1) that we will need for our work below. In operator form, Equation (1) has the solution $\rho(t) = \wp(t)\rho(0)$, where the propagator $\wp(t)$ is given by [18]

$$\wp(t) = \exp\left[-\frac{w_4}{e^{w_4}-1}\left(w_1\{q,\cdot,q\} + w_2\{p,\cdot,p\} - \left(w_3 - i\frac{e^{w_4}-1}{4\hbar}\right)\{p,\cdot,q\} - \left(w_3 + i\frac{e^{w_4}-1}{4\hbar}\right)\{q,\cdot,p\}\right)\right](U_s \cdot U_s^\dagger) \quad (3)$$

where $U_s$ satisfies $\frac{dU_s(t)}{dt} = \frac{1}{\hbar i}H_s(t)U_s(t)$ with initial condition $U_s(0)=1$, and where the coefficients in the preceding exponent are solutions to the following inhomogeneous system of differential equations:

$$\frac{d}{dt}\begin{pmatrix} w_1 \\ w_2 \\ w_3 \end{pmatrix} = 2\begin{pmatrix} -2b_{12} & 0 & -2b_{11} \\ 0 & 2b_{12} & 2b_{22} \\ b_{22} & -b_{11} & 0 \end{pmatrix}\begin{pmatrix} w_1 \\ w_2 \\ w_3 \end{pmatrix} + e^{w_4}\begin{pmatrix} k_1 \\ k_2 \\ (k_3 + k_4)/2 \end{pmatrix} \quad (4)$$

with $w_4 = 2\hbar i \int_0^t (k_3 - k_4)dt'$, and initial conditions $w_1(0) = w_2(0) = w_3(0) = 0$. Not all $b_{ij}$ and $k_i$ give rise to physical solutions; to ensure that $\rho$ is normalized, Hermitian and positive, we shall assume that these coefficients yield $w_1 > 0$, $w_2 > 0$, $w_1 w_2 - w_3^2 - \frac{(e^{w_4}-1)^2}{16\hbar^2} \geq 0$ and positive dissipation, $w_4 > 0$, for $t > 0$ [18].

## 3. Noise Reduction in Quantum Dissipative Equations

A state evolving under a generic QDE loses coherence as pure states map to mixed states. Preferred states that minimize entropy production have been investigated for some master equations [26,27]. Better, we show that under a simple transformation, which we dub the deconvolution picture (DP), we can eliminate decoherence altogether: for any QDE, there exists a pair of time-dependent generalized lowering operators, $C(t)$ and $B(t)$, obeying commutation relations $[C(t), C^\dagger(t)] = [B(t), B^\dagger(t)] = 1$ and having squeezed eigenstates and complex eigenvalues satisfying

$$C(t)|\gamma\rangle_{C(t)} = \gamma|\gamma\rangle_{C(t)} \quad (5)$$

and

$$B(t)|\beta\rangle_{B(t)} = \beta|\beta\rangle_{B(t)}, \quad (6)$$

such that every eigenstate of $C(t^*)$ evolves after an arbitrary but fixed time $t^*$ to an eigenstate of $B(t^*)$ in the DP. Above and throughout this paper, we use Dirac ket notation $|\psi\rangle$ to represent a state vector $\psi$. In addition to obeying the foregoing commutation relation, generalized lowering and raising operators are linear combinations, with complex coefficients, of the usual lowering and raising (or position and momentum) operators, and have been extensively studied in the literature [28].

In one example, the DP (denoted by a superscript *D*) is defined by

$$\rho^D(t) = \exp[\delta(t)\{p,\cdot,p\}]\rho(t), \quad (7)$$

where $\rho(t)$ is a density operator obeying a QDE, and $\delta(t)$ is a non-negative *c*-function with initial condition $\delta(0) = 0$ to be specified below (we will see that more generally, the DP may be defined via a real, linear combination of $\{q,\cdot,q\}$ and $\{p,\cdot,p\}$). The operator $\exp[\delta(t)\{p,\cdot,p\}]$ corresponds to the aforementioned impurity filter converting states that are not pure to squeezed states that are, but only at a privileged time $t^*$, which can act as a key for secure communication.

If there were a time when $\delta(t)$ would vanish, the deconvolution and Schrodinger pictures would coincide and, like in Equation (2) at absolute zero, only one dissipative operator, $\{B^\dagger, \cdot, B^\dagger\}$, would come into play.

For a generic QDE, $\delta(t)$ is typically not zero for $t > 0$. Nevertheless, although $\delta(t)$ may not vanish, we can propagate wave functions instead of density operators in the DP, as with Hasse's non-linear Schrodinger equation. Assuming that the initial state is $\rho(0) = |\beta\rangle_{C(t^*)C(t^*)}\langle\beta|$, we will see below that the usual Schrodinger picture may be obtained at $t^*$ via $\rho(t^*) = \exp[-\delta(t^*)\{p,\cdot,p\}]|\beta e^{-w_4(t^*)/2}\rangle_{B(t^*)B(t^*)}\langle\beta e^{-w_4(t^*)/2}|$. When we compare this expression with Equation (7), we see that $\rho^D(t^*) = |\beta e^{-w_4(t^*)/2}\rangle_{B(t^*)B(t^*)}\langle\beta e^{-w_4(t^*)/2}|$; i.e., under any QDE, we can find a time $t^*$ and generalized lowering operators $B(t)$ and $C(t)$, such that to within a global phase factor, the initial pure state $|\beta\rangle_{C(t^*)}$ evolves after $t^*$ to another pure state $|\beta e^{-w_4(t^*)/2}\rangle_{B(t^*)}$ in the DP. Having described in general terms the role of the operators $B(t)$ and $C(t)$, we now characterize them in more detail.

The generalized lowering operator $B$ [29] is defined by

$$B(t) = \sqrt{\frac{2}{e^{w_4}-1}}\left(e^{-i\xi}\sqrt{w_1-r_1}\,q + e^{i\xi}\sqrt{w_2-r_2}\,p\right) \tag{8}$$

where $r_1(t)$ and $r_2(t)$ are any non-negative functions that at $t=0$ vanish and at later times satisfy $w_1 - r_1 > 0$, $w_2 - r_2 > 0$ and $(w_1-r_1)(w_2-r_2) = w_3^2 + \frac{(e^{w_4}-1)^2}{16\hbar^2}$, and where

$$0 \le \xi = \frac{1}{2}\tan^{-1}\left(-\frac{e^{w_4}-1}{4\hbar w_3}\right) \le \frac{\pi}{2}. \tag{9}$$

This generalized lowering operator satisfies the eigenvalue Equation (6). The propagator given by Equation (3) can be factored (see [18] for a related factorization) so that the only operator appearing in the non-Hamiltonian part in the DP is $B^\dagger$:

$$\rho^D(t) = \exp\left(-\frac{w_4}{2}\{B^\dagger,\cdot,B^\dagger\}\right)\left(U_s\rho(0)U_s^\dagger\right) \tag{10}$$

where the DP is defined by

$$\rho^D(t) = \exp\left[e^{-w_4}(r_1\{q,\cdot,q\} + r_2\{p,\cdot,p\})\right]\rho(t). \tag{11}$$

The operator $\exp[e^{-w_4}(r_1\{q,\cdot,q\} + r_2\{p,\cdot,p\})]$ is the general form of the impurity filter. Because of the freedom in choosing $r_1$ and $r_2$, the factorization (10) is not unique. In Equation (7), we chose the impurity filter $\exp[\delta(t)\{p,\cdot,p\}]$ corresponding to $r_1 = 0$ and $\delta(t) = e^{-w_4}r_2$. The generalized lowering operator $C(t)$ is generally defined by

$$C(t) = U_s^\dagger(t)B(t)U_s(t), \tag{12}$$

and satisfies the eigenvalue Equation (5). Letting the initial state be $\rho(0) = |\beta\rangle_{C(t^*)C(t^*)}\langle\beta|$, where $t^*$ is an arbitrary but fixed time and $\beta$ is variable, and using the general formula

$$e^{-f\{B^\dagger,\cdot,B^\dagger\}}|\beta\rangle_B {}_B\langle\beta| = |\beta e^{-f}\rangle_B {}_B\langle\beta e^{-f}| \qquad (13)$$

for $f \geq 0$ and $B$ any generalized lowering operator, we obtain at time $t^*$

$$\rho^D(t^*) = |\beta e^{-w_4(t^*)/2}\rangle_{B(t^*)} {}_{B(t^*)}\langle\beta e^{-w_4(t^*)/2}|. \qquad (14)$$

With this evolution, we can associate an operator $\hat{W}_{DP}(t)$ that propagates kets, $|\phi(t)\rangle = \hat{W}_{DP}(t)|\phi(0)\rangle$, defined by

$$\hat{W}_{DP}(t)|\phi(0)\rangle = \frac{\exp\left(-\frac{w_4(t)}{2}B^\dagger(t)B(t)\right)U_s(t)|\phi(0)\rangle}{\left\|\exp\left(-\frac{w_4(t)}{2}B^\dagger(t)B(t)\right)U_s(t)|\phi(0)\rangle\right\|}. \qquad (15)$$

For any QDE, the density operator corresponding to the state $|\phi(0)\rangle = |\beta\rangle_{C(t^*)}$, with $\beta$ variable, evolves in the DP to the density operator corresponding to the state $|\phi(t^*)\rangle = \hat{W}_{DP}(t^*)|\phi(0)\rangle = |\beta e^{-w_4(t^*)/2}\rangle_{B(t^*)}$.

## 4. An Example Using the Wehrl Entropy

To further elucidate our work and characterize the privileged time $t^*$, we employ the Wehrl entropy of a density operator $\rho$, denoted by $S_W(\rho)$. The quasiclassical quantity $S_W(\rho)$ was introduced by Wehrl [30], and is defined according to

$$S_W(\rho) = -\frac{1}{2\pi}\int d\tilde{x}\int d\tilde{p}\, {}_a\langle\alpha|\rho|\alpha\rangle_a \ln {}_a\langle\alpha|\rho|\alpha\rangle_a \qquad (16)$$

where $|\alpha\rangle_a$ are the well-known coherent eigenstates of the lowering operator $a$, which in connection with a harmonic oscillator of mass $m$ and frequency $\omega$ are given in the position representation by

$$\langle x | \alpha\rangle_a = \left(\frac{m\omega}{\pi\hbar}\right)^{1/4} \exp\left[-\frac{1}{2}\left(\sqrt{\frac{m\omega}{\hbar}}x - \tilde{x}\right)^2 + i\sqrt{\frac{m\omega}{\hbar}}x\tilde{p}\right] \qquad (17)$$

with $\tilde{x}$ and $\tilde{p}$ real, and $\alpha = (\tilde{x} + i\tilde{p})/\sqrt{2}$. In the rest of this section, we will set $\hbar = \omega = m = 1$. Given the von Neuman entropy $S(\rho) = -Tr\rho\ln\rho$, where Boltzmann's constant has been set to unity, we have the following two inequalities, the first of which was proven in [30], and the second conjectured in [30] and proven by Lieb [31]:

$$S_W(\rho) > S(\rho) \qquad (18)$$

and

$$S_W(\rho) \geq 1, \qquad (19)$$

where this last expression becomes an equality if $\rho = |\alpha\rangle_a {}_a\langle\alpha|$.

We investigate a simple example, in which $B$ is time independent and given by

$$B = e^{-i\pi/4} a \tag{20}$$

where $a = (q + ip)/\sqrt{2}$ is the usual lowering operator. We note that the operator $e^{-i\pi/4} a$ has coherent eigenstates $|\alpha\rangle_a$ with eigenvalue $e^{-i\pi/4}\alpha$. Thus, $|\alpha\rangle_a = |e^{-i\pi/4}\alpha\rangle_{e^{-i\pi/4}a}$. Comparing to Equation (8), this implies $\xi = \pi/4$,

$$w_1 - r_1 = \frac{e^{w_4} - 1}{4} \tag{21}$$

$$w_2 - r_2 = \frac{e^{w_4} - 1}{4} \tag{22}$$

and

$$w_3 = 0. \tag{23}$$

Using these special values in Equation (10) and noting that $\{e^{i\pi/4}a^\dagger, \cdot, e^{i\pi/4}a^\dagger\} = \{a^\dagger, \cdot, a^\dagger\}$, the density operator in the DP becomes

$$\rho^D = \exp\left(-\frac{w_4}{2}\{a^\dagger, \cdot, a^\dagger\}\right)\left(U_s^{-1}(t^*, t) e^{i\pi/4}\beta\right)_{a\,a}\left\langle e^{i\pi/4}\beta \big| U_s(t^*, t)\right). \tag{24}$$

The squeezed state

$$U_s^{-1}(t^*, t)\big|e^{i\pi/4}\beta\big\rangle_a = U_s^{-1}(t^*, t)\big|\beta\big\rangle_{e^{-i\pi/4}a} \tag{25}$$

$$= \big|\beta\big\rangle_{A(t^*, t)} \tag{26}$$

is an eigenstate of the generalized lowering operator $A(t^*, t) = U_s^{-1}(t^*, t) e^{-i\pi/4} a U_s(t^*, t) = \mu(t^*, t) a + \nu(t^*, t) a^\dagger$ with eigenvalue $\beta$. Because $[A(t^*, t), A^\dagger(t^*, t)] = 1$, we have $|\mu|^2 - |\nu|^2 = 1$ (see Yuen [28]). These relations imply that $\mu(t^*, t^*) = e^{-i\pi/4}$ and $\nu(t^*, t^*) = 0$.

The foregoing relations impose restrictions on the QDE that is supposed to yield solution (24) and it is helpful to summarize what these are. Our initial time is taken to be zero and the privileged time satisfies $t^* > 0$. We choose the Hamiltonian variables $b_{11}(t^*, t)$, $b_{12}(t^*, t)$ and $b_{22}(t^*, t)$ such that $\mu(t^*, t^*) = e^{-i\pi/4}$ and $\nu(t^*, t^*) = 0$. Next, we choose $k_1$, $k_2$ and $k_3$ such that $w_1 > (e^{w_4} - 1)/4$, $w_2 > (e^{w_4} - 1)/4$ and $w_3 = 0$ for $t > 0$. We finally choose $r_1 = w_1 - (e^{w_4} - 1)/4$ and $r_2 = w_2 - (e^{w_4} - 1)/4$.

It is useful to introduce a parameter $\tau \geq 0$ into Equation (24), which parameter we will later set equal to unity, so that

$$\rho^D(\tau) = \exp\left(-\frac{w_4 \tau}{2}\{a^\dagger, \cdot, a^\dagger\}\right)\left(U_s^{-1}(t^*, t) e^{i\pi/4}\beta\right)_{a\,a}\left\langle e^{i\pi/4}\beta \big| U_s(t^*, t)\right). \tag{27}$$

Differentiating with respect to $\tau$ yields the differential equation

$$\frac{d\rho^D(\tau)}{d\tau} = -\frac{w_4}{2}\{a^\dagger, \rho^D(\tau), a^\dagger\} \tag{28}$$

with a squeezed state "initial condition"

$$\rho^D(0) = U_s^{-1}(t^*,t)\left|e^{i\pi/4}\beta\right\rangle_{a\,a}\!\left\langle e^{i\pi/4}\beta\right|U_s(t^*,t). \tag{29}$$

Conveniently, a solution of Equation (28) with an initial squeezed state has been computed in the literature [32]. That solution, applied to our particular values, is

$$_a\langle\alpha|\rho^D(\tau)|\alpha\rangle_a = \frac{1}{\sqrt{|\det\sigma(\tau)|}}\exp\!\left[-\frac{1}{2}\bigl(\mathbf{z}-\langle\mathbf{z}\rangle e^{-w_4\tau/2}\bigr)^T\sigma^{-1}(\tau)\bigl(\mathbf{z}-\langle\mathbf{z}\rangle e^{-w_4\tau/2}\bigr)\right] \tag{30}$$

where

$$\sigma(\tau) = \begin{pmatrix} -\mu^*\nu e^{-w_4\tau} & 1+e^{-w_4\tau}|\nu|^2 \\ 1+e^{-w_4\tau}|\nu|^2 & -\mu\nu^* e^{-w_4\tau} \end{pmatrix}, \tag{31}$$

$$\mathbf{z} = \begin{pmatrix} \alpha \\ \alpha^* \end{pmatrix} \tag{32}$$

and

$$\langle\mathbf{z}\rangle = \begin{pmatrix} {}_a\langle\beta|a|\beta\rangle_a \\ {}_a\langle\beta|a^\dagger|\beta\rangle_a \end{pmatrix}. \tag{33}$$

We can use Equation (30) in the definition of the entropy (16), and consult a table of multivariate Gaussian integrals to compute the Wehrl entropy of the density operator in the DP. The parameter $\tau$ was introduced merely as an aid for computation, and we ultimately set it to unity to obtain

$$S_W(\rho^D) = 1 + \ln\sqrt{1+|\nu|^2 e^{-w_4}(2-e^{-w_4})}, \tag{34}$$

where we note that for the subset of squeezed states characterized by one squeezing parameter, this last Wehrl entropy appears in Reference [33].

As a check of our work, we note that when we take the noise free case with $w_4 = 0$, the system evolves as a time dependent squeezed state. In such a case, Equation (34) predicts a value for the Wehrl entropy of $1+\ln\sqrt{1+|\nu|^2}$, which agrees with the value for a squeezed state found in the literature [34]. For the physically reasonable choice of positive dissipation, for which $w_4 \geq 0$ for $t \geq 0$, we see that $S_W(\rho^D)$ is greater than its minimum value of unity unless $|\nu| = 0$. Because $\nu(t^*,t^*) = 0$, the Wehrl entropy at the privileged time $t^*$ is unity and $\rho^D$ is then a coherent state that may be found with the help of Equation (13):

$$\rho^D(t^*) = e^{-\frac{w_4}{2}\{a^\dagger,\cdot,a^\dagger\}}\left(\left|e^{i\pi/4}\beta\right\rangle_{a\,a}\!\left\langle e^{i\pi/4}\beta\right|\right) \tag{35}$$

$$= \left|\beta e^{-\frac{1}{2}(w_4-i\pi/2)}\right\rangle_{a\,a}\!\left\langle \beta e^{-\frac{1}{2}(w_4-i\pi/2)}\right|. \tag{36}$$

This result is consistent with the observation in Reference [24] that for the rotating wave optical master equation (Equation (2)) at absolute zero, a coherent state remains a coherent state.

To sum up, for the example treated in this section in the DP, we start off at time $t=0$ in a squeezed eigenstate of the operator $A(t^*,0)$, and after a time $t^*$, we end up in the coherent state given by Equation (36).

If we take the Wehrl entropy as a measure of noise, we can provide a rough estimate of the time interval during which the noise will be less than some tolerance. In other words, we seek an interval $|t - t^*| \leq \Delta t$ so that $S_W(\rho^D, t^*, t) - 1 \leq \varepsilon$. For small $\Delta t$, we can Taylor expand $S_W(\rho^D, t^*, t) - 1$ as a function of time about $t^*$. Assuming $|v(t)|$ is sufficiently smooth with a minimum at $t = t^*$ so that $|v(t^*, t^*)| = \frac{\partial |v(t, t^*)|}{\partial t}\bigg|_{t=t^*} = 0$, we have to go out to fourth order before we find a non-vanishing term. Ignoring higher order terms, we find $S_W(\rho^D, t^*, t) - 1 \approx \frac{1}{8} e^{-w_4(t^*)} \left(2 - e^{-w_4(t^*)}\right) \left(\frac{\partial^2 |v(t)|}{\partial t^2}\bigg|_{t=t^*}\right)^2 (t - t^*)^4$, whence we obtain

$$\Delta t \approx \left(8\varepsilon \frac{e^{2w_4(t^*)}}{2e^{w_4(t^*)} - 1}\right)^{1/4} \left(\frac{\partial^2 |v(t)|}{\partial t^2}\bigg|_{t=t^*}\right)^{-1/2}. \tag{37}$$

The time $\Delta t$ ultimately depends on some of the parameters of the underlying QDE through the dependence of $w_4$ on the difference $k_3 - k_4$, via $w_4 = 2\hbar i \int_0^t (k_3 - k_4) dt'$, and the dependence of $|v|$ on the Hamiltonian parameters $b_{11}$, $b_{12}$ and $b_{22}$, via $U_s^{-1}(t^*, t) e^{-i\pi/4} a U_s(t^*, t) = \mu(t^*, t) a + v(t^*, t) a^\dagger$.

## 5. Coherent Processing of a Qubit

A qubit, approximated as a superposition of substantially non-overlapping coherent states, is prone to decoherence. Thus, we avoid such a superposition and instead represent a qubit by a single squeezed state whose position in phase space encodes information. The impurity filter enables processing without decoherence. In the following formalism, unitary operators that keep the set of squeezed coherent states invariant are candidates for single qubit gates.

We adopt the observable $P_1 = \int_0^\infty dq |q\rangle\langle q|$, noting that similar projectors arise in the study of insufficiently selective measurements [35] and homodyne detection [36]. The observable $P_1$ has two eigenvalues 0 and 1, each having an uncountably infinite degree of degeneracy. Given a squeezed state $|\beta\rangle_B$, we next define $|0; \beta\rangle_B = P_0 |\beta\rangle_B = (1 - P_1)|\beta\rangle_B$ and $|1; \beta\rangle = P_1 |\beta\rangle_B$. According to convention [1], a qubit $|\psi\rangle$ is a vector in a two-dimensional state space that is given by

$$|\psi\rangle = a|0\rangle + b|1\rangle, \tag{38}$$

where the kets $|0\rangle$ and $|1\rangle$, constituting an orthonormal basis, are fixed and the coefficients, satisfying the normalization condition $|a|^2 + |b|^2 = 1$, are otherwise variable. We may alternatively represent a qubit by a squeezed state $|\beta\rangle_B$, which can be decomposed as

$$|\beta\rangle_B = |0; \beta\rangle_B + |1; \beta\rangle_B, \tag{39}$$

where now, opposite to the conventional formalism, the last two kets can vary by changing $\beta$ and the coefficients are fixed to unity.

States differing by only a global phase factor are assumed to be equivalent. Therefore, only two parameters, $\theta$ and $\phi$, are needed to specify $a$ and $b$, which then fix the qubit $|\psi\rangle$ [1]. We may parametrize the qubit, as follows:

$$|\psi\rangle = \frac{1}{\sqrt{2}} \text{erfc}^{1/2}(\theta)|0\rangle + \frac{1}{\sqrt{2}} e^{i\phi} \text{erfc}^{1/2}(-\theta)|1\rangle, \tag{40}$$

where $\text{erfc}(x)$ is the complementary error function. For fixed variances and to within a global phase factor, a squeezed state is in one-to-one correspondence with its two first moments, $\langle q \rangle$ and $\langle p \rangle$. Thus, with the help of the equations

$$\theta = \langle q \rangle / (\sqrt{2}\Delta q) \tag{41}$$

and

$$\phi = \langle p \rangle / \Delta p, \tag{42}$$

where $\Delta q$ and $\Delta p$ are variances of this state, we can uniquely map a squeezed state $|\beta\rangle_B$ (or, if the variances are held fixed, its first two moments) to a qubit $|\psi\rangle$. The probability of measuring a result of 0 or 1 for the qubit $|\psi\rangle$ is $\text{erfc}(\theta)/2$ and $\text{erfc}(-\theta)/2$, respectively. These are the same probabilities, ${}_B\langle\beta|0;\beta\rangle_B$ and ${}_B\langle\beta|1;\beta\rangle_B$, that a measurement of the observable $P_1$ would yield the eigenvalues 0 or 1 in the state $|\beta\rangle_B$, which explains why we chose the foregoing parametrization for the qubit.

In view of this correspondence between squeezed states and qubits, the $X$ (or *NOT*), $Y$ and $Z$ gates [1] may be implemented with the operators $\Pi$, $T(0, \pi\Delta p)\Pi$ and $T(0, \pi\Delta p)$, respectively, where $\Pi$ is the parity operator defined by $\Pi|q\rangle = |-q\rangle$ [35], and $T(q_0, p_0)$ is the translation operator that is given by $\exp[i(q_0 p - p_0 q)/\hbar]$, where $q_0$ and $p_0$ are real parameters. This is consistent with the observation that $ZX$ and $Y$, related by $ZX = e^{i\pi/2} Y$, yield the same qubits to within a global phase factor.

With the aid of an impurity filter, it is possible to perform coherent quantum computation on single qubits evolving under a QDE. Let us look at the following quantum circuit for the *NOT* gate as an illustrative example:

$$|\beta\rangle_{B(t^*)} \xrightarrow{\hat{W}_{DP}(t^*) U_s^{-1}(t^*)} \left|\beta e^{-w_4(t^*)/2}\right\rangle_{B(t^*)} \boxed{NOT} \left|-\beta e^{-w_4(t^*)/2}\right\rangle_{B(t^*)}. \tag{43}$$

This circuit may be implemented with QDE dynamics as follows:

$$\Pi\left[e^{\delta(t^*)\{p,\cdot,p\}}\wp(t^*)\left[U_s^{-1}(t^*)|\beta\rangle_{B(t^*)B(t^*)}\langle\beta|U_s(t^*)\right]\right]\Pi^{-1} = \left|-\beta e^{-w_4(t^*)/2}\right\rangle_{B(t^*)B(t^*)}\left\langle-\beta e^{-w_4(t^*)/2}\right| \tag{44}$$

After choosing a time $t^*$, we prime the input state $|\beta\rangle_{B(t^*)B(t^*)}\langle\beta|$ to $U_s^{-1}(t^*)|\beta\rangle_{B(t^*)B(t^*)}\langle\beta|U_s(t^*)$. Next, we let QDE dynamics run for $t^*$. We then apply the impurity filter and finally the parity operator to invoke the *NOT* gate. The result is the state $\left|-\beta e^{-w_4(t^*)/2}\right\rangle_{B(t^*)B(t^*)}\left\langle-\beta e^{-w_4(t^*)/2}\right|$.

Turning to two qubits, we may take the tensor product of two squeezed states, $|\alpha\rangle_B \otimes |\beta\rangle_B$, as the fiducial input. However, because gates for multiple qubits generally convert product states to correlated states, we must admit entangled states into the formalism. A *CNOT* gate [1,10] serves as an example. Conventionally, this gate may be written as

$$\tilde{P}_0 \otimes 1 + \tilde{P}_1 \otimes X \tag{45}$$

where $\tilde{P}_j \equiv |j\rangle\langle j|$, with $j = 0$ or 1. Similarly, in the above formalism where the parity operator corresponds to the $X$ gate, we can represent the *CNOT* gate, as follows:

$$P_0 \otimes 1 + P_1 \otimes \Pi. \tag{46}$$

For example, taking as input the state $|0\rangle_B \otimes |\beta\rangle_B$ and making use of the standard expression for a squeezed state [28] yields the following output in the position representation:

$$\begin{aligned}&(\langle q_1| \otimes \langle q_2|)CNOT(|0\rangle_B \otimes |\beta\rangle_B) \\ &= \left(\frac{m\omega}{\pi\hbar}\right)^{1/4} (\mu - \nu)^{-1/2} \exp\left(-\frac{\mu+\nu}{\mu-\nu}\frac{m\omega}{2\hbar}q_1^2\right)\left[\theta(-q_1)\langle q_2 | \beta\rangle_B + \theta(q_1)\langle q_2 | -\beta\rangle_B\right]\end{aligned} \tag{47}$$

where $\theta(q)$ is the unit step function and $B = \mu a + \nu a^\dagger$.

## 6. Discussion

It is tempting to think of the impurity filter given by Equation (7) as the inverse of $e^{-\delta\{p,\cdot,p\}}$, but this would be strictly incorrect because $e^{-\delta\{p,\cdot,p\}}\left(e^{\delta\{p,\cdot,p\}}\rho\right)$ is not defined for all density operators $\rho$ (recall that $\delta \geq 0$). For example, for certain bone fide $\rho$, the operator $e^{\delta\{p,\cdot,p\}}\rho$ may not be positive. Worse, $e^{\delta\{p,\cdot,p\}}\rho$ may not even exist. To see this, formally express $e^{\delta\{p,\cdot,p\}}\rho$ in the Wigner representation: $\left[e^{\delta\{p,\cdot,p\}}\rho\right]_W = \frac{1}{2\sqrt{\pi\delta}}\int_{-\infty}^{\infty}\exp\left(-\frac{x^2}{4\delta}\right)W(p, q - i\hbar x)dx$ where $W$ is the Wigner function corresponding to $\rho$ [37]. For the state $\rho = |0\rangle_{BB}\langle 0|$, for instance, we require that $\delta < \frac{1}{8\langle p^2\rangle}$ for the last integral to exist. More insight can be gained by looking at $\left[e^{-\delta\{p,\cdot,p\}}\rho\right]_W = \frac{1}{2\sqrt{\pi\delta}}\int_{-\infty}^{\infty}\exp\left(-\frac{x^2}{4\delta}\right)W(p, q - \hbar x)dx$. This is a Gaussian convolution in the position variable, which explains our choice of the phrase "deconvolution picture" when applying the operator $e^{\delta\{p,\cdot,p\}}$.

Unlike a typical propagator corresponding to Hasse's non-linear Schrodinger equation, the wave-function propagator $\hat{W}_{DP}(t)$ of Equation (15) does not describe the evolution of the related master equation for all times and initial states. Rather, for any arbitrary but fixed time $t^*$, and for any QDE, we have seen that there is a squeezed state basis that, in the DP, evolves after the time $t^*$ to another squeezed state basis according to this last propagator; before or after such time, the states need not be pure.

Quantum communication is hampered by a no-go theorem of Niset et al.: Gaussian operations cannot shield states from errors when these states and errors are also Gaussian [38]. A lot of work in the literature directed to correcting errors arising from quantum noise makes use of redundancy of states. In contrast, the work herein exploits a unique factorization of the propagator, followed by a deconvolution that leaves squeezed states coherent and accessible to quantum processing. A significance of this approach is that quantum noise is attenuated without having to increase the dimensions of the system under study, albeit at the expense of having to perform a deconvolution at a privileged time $t^*$. This limitation could be used advantageously for secure information transfer. An eavesdropper applying an impurity filter at an arbitrary time will not generally recover a pure state except in the unlikely event that he or she happens to apply the filter at precisely $t^*$. An authorized person would be apprised of this time, and therefore could recover such a state.